# Integrating Flowsheet Data in OMOP Common Data Model for Clinical Research

Tina Seto[1], Lillian Sung[2], Jose Posada[3], Priyamvada Desai[1], Susan Weber[1], Somalee Datta[1]

1: Technology & Digital Solutions, Stanford Medicine; 2: Division of Haematology Oncology, The Hospital for Sick Children, Toronto, Ontario, Canada; 3: Center for Biomedical Informatics Research, Stanford University

## Abstract

Flowsheet data presents unique challenges and opportunities for integration into standardized Common Data Models (CDMs) such as the Observational Medical Outcomes Partnership (OMOP) CDM from the Observational Health Data Sciences and Informatics (OHDSI) program. These data are a potentially rich source of detailed curated health outcomes data such as pain scores, vital signs, lines drains and airways (LDA) and other measurements that can be invaluable in building a robust model of a patient's health journey during an inpatient stay. We present two approaches to integration of flowsheet measures into the OMOP CDM. One approach was computationally straightforward but of potentially limited research utility. The second approach was far more computationally and labor intensive and involved mapping to standardized terms in controlled clinical vocabularies such as Logical Observation Identifiers Names and Codes (LOINC), resulting in a research data set of higher utility to population health studies.

## Introduction

Flowsheet data in electronic health records (EHR) contain a wealth of information that may be very useful for researchers looking for data beyond diagnosis codes, procedures, labs, encounters, medications, notes and demographics [Johnson2015]. Unlike the coding for diagnoses, procedures and drugs which typically use universal standard terminologies to facilitate billing, flowsheet measure names and values may lack standardization. There may be multiple terms with different row names yet are semantically equivalent such as temperature, Temp, tcore, Temp in Celsius. At the same time, row names may be case-sensitive in that lower-case 'temp' would not be referencing temperature. Flowsheet data may also be recorded in different units, even within the same hospital. In addition to challenges in normalizing the data [Westra2015], many data engineers and clinical researchers have little idea what data are available in flowsheets. Flowsheet templates are often customized to the workflow of the institution or even a department within the institution, resulting in wide variations in labeling, nomenclature, abbreviations and units of measurement. Aside from the complexity of the content, the data engineer will also likely be faced with the operational challenge of processing and manipulating a set of tables in a database with billions of rows that may grow exponentially as more flowsheet templates and questionnaires are added. Lastly, the data engineer needs to decide which of the tens of thousands of flowsheet row measures would be most useful to include in a common data model (CDM) for the researchers at their own institution as well as other institutions if participating in network studies.

# Background

In Epic, flowsheets are comprised of templates, groups and rows. Rows contain the patient id, recorded time, provider id and measured value which can be numeric, a selected value from a custom list or free text. Typically, flowsheet data are entered by nurses, medical assistants, nursing assistants, respiratory therapists, physical therapists and case managers to document assessments, observations, interventions, checklists for routine care tasks, anesthesia events, line placements, surgical time events, patient goals, social determinants of health and other longitudinal data [Winden2018]. Flowsheet templates are customized to each institution and created as needed to streamline and standardize workflow. Often, there are multiple templates capturing the similar data as different departments have specific requirements. Very few of the over twenty thousand row measure types are coded to standard terminologies such as Logical Observation Identifiers Names and Codes (LOINC) or Systematized Nomenclature of Medicine - Clinical Terms (SNOMED-CT). This vast, unorganized data store creates a secondary-use challenge [Westra2017] to data analysts and informaticians on how to find what flowsheet data to extract and how to represent that data uniformly in network studies and publications. Since 2013, the School of Nursing at the University of Minnesota has been hosting the Nursing Knowledge Big Data Science Conference to help create common information models and standardizing nurse-sensitive data [Waitman2011]. Even if the various electronic health records (EHR) software vendors integrate new common models into their systems, this would mainly benefit new customers. Most of the large health organizations in the US transitioned from paper or primitive EHR software to modern EHR systems in the 2000s and are already established in their customized nursing and other allied health data workflow.

# Data source

The STAnford medicine Research data Repository (STARR) includes an OMOP-formatted clinical data warehouse (CDW) named STARR-OMOP. It contains EHR data from the two hospitals in Stanford Health Care (SHC) and Stanford Children's Health (SCH), as well as more than 60 clinics and centers across the peninsula, ?and East bay and South bay in the San Francisco bay area [Datta2015] . STARR-OMOP is maintained by Research IT in the Technology and Digital Solutions (TDS) department within Stanford Health Care. As of June 2021, STARR-OMOP represents nearly 3.3 million patients and 6.8 million data points that include encounters, labs, medications, procedures, diagnoses, notes, flowsheet, and Admission/Discharge/Transfer (ADT) records.

# Methods

The integration of flowsheet data in STARR-OMOP was a two-step process. The first version of STARR-OMOP used a computationally simple approach to make all flowsheet data available to researchers who had IRB and Privacy Office permission to view Protected Health Information (PHI) data.  With 1,883 templates, 46,85 groups and 25,777 distinct types of row measures, manual mapping to standard terminology would have been too time-consuming.  Software tools have been created by other institutions to map flowsheet data such as FloMap developed at the University of Minnesota [Johnson2017] or machine learning models [Johnson2020] to automate the mapping but these tools are not open-sourced or productized.  There have been various approaches to standardize and create flowsheet ontologies [Harris2015], [Kim2010], [Lytle2021], [Warren2012]. To make all flowsheet data readily available and as quickly as possible, we created JavaScript Object Notation (JSON) strings for each flowsheet row name and value pair.  This added 3.6 billion rows (52%) to the OBSERVATION table and enabled research use of unprocessed data. The JSON strings included flowsheet values containing free text such as names, phone numbers, addresses and possibly other PHI data.  For researchers without a PHI-permissible IRB, the flowsheet JSON strings were censored in its entirety from the anonymized version of our OMOP.

This method comprehensively added all flowsheets, thereby allowing research of all types. However, the unprocessed nature limited broad utility for population health studies that tend to rely on curated mappings to standardized terminologies. Furthermore, some flowsheet row types were quality-improvement related and of little interest to researchers such as the order in which the bedrails were lifted and name of the driver picking up at discharge. The STARR-OMOP database is refreshed weekly, and unnecessary/irrelevant data incrementally added to the cost of the CDM with no recognizable return on investment.

In the second stage of the project, we selectively mapped high-value flowsheet entries to standardized terminologies. In support of a clinical investigator interested in a network study involving bloodstream infections (BSI) in pediatric cancer patients [Sung2020], we mapped 26 flowsheet measures to standard LOINC concept codes as described in Table 1. The investigator along with members of the Center for Biomedical Informatics Research (BMIR) used manual mapping techniques to arrive at a consensus for the mapping between the flowsheet row name and LOINC code. The mapping was then provided to a data engineer within Research IT to implement into the OMOP data transformation pipeline. To facilitate de-identifying our OMOP, we only mapped flowsheet concepts with numeric measure values for this initial effort.

Table 1: Mapping of flowsheets to standard LOINC concept codes

| concept_id | concept_code | vocabulary_id | concept_name | common abbr |
| --- | --- | --- | --- | --- |
| 3005424 | 8277-6 | LOINC | Body surface area | BSA |
| 3020891 | 8310-5 | LOINC | Body temperature | Temp, Tcore, Temperature |
| 3025315 | 29463-7 | LOINC | Body weight | Weight |
| 21490675 | 60985-9 | LOINC | Central venous pressure (CVP) | CVP |
| 3012888 | 8462-4 | LOINC | Diastolic blood pressure | Diastolic BP, ARTD, Arterial Diastolic BP |
| 21490565 | 60802-6 | LOINC | Dynamic plateau pressure | Pplat |
| 3032652 | 35088-4 | LOINC | Glasgow coma scale | Glasgow |
| 3027018 | 8867-4 | LOINC | Heart rate | HR |
| 3036277 | 8302-2 | LOINC | Height | Height |
| 3005629 | 3151-8 | LOINC | Inhaled oxygen flow rate | O2 |
| 45876241 | IO_OUT | LOINC | Input/Output | Urine, Urine Output |
| 21490581 | 60826-5 | LOINC | Lung compliance | COMP |
| 42527086 | 60949-5 | LOINC | Mean airway pressure | MnAwP |
| 3027598 | 8478-0 | LOINC | Mean blood pressure | ARTM, Mean Arterial Pressure |
| 21490566 | 60804-2 | LOINC | Minimum alveolar concentration (MAC) for anesthesia | etMAC |
| 3045410 | 33425-0 | LOINC | Minute volume setting Ventilator | MV |
| 21490615 | 60860-4 | LOINC | Nitrous oxide [VFr/PPres] Gas delivery system | N2O |
| 3024882 | 19994-3 | LOINC | Oxygen/Inspired gas setting [Volume Fraction] Ventilator | FiO2 |
| 21490855 | 76248-4 | LOINC | PEEP Respiratory system --on ventilator | PEEP |
| 3036453 | 38214-3 | LOINC | Pain severity [Score] Visual analog score | Pain Level |
| 3011557 | 19931-5 | LOINC | Peak inspiratory gas flow setting Ventilator | PIP |
| 3025809 | 8634-8 | LOINC | Q-T interval | QT Interval |
| 3026258 | 8636-3 | LOINC | Q-T interval corrected | QTc Interval |
| 3024171 | 9279-1 | LOINC | Respiratory rate | Resp, Resp rate |
| 21490553 | 60782-0 | LOINC | Sevoflurane gas delivered during case [Volume] from Gas delivery system | Sevoflurane |
| 3004249 | 8480-6 | LOINC | Systolic blood pressure | Systolic BP, ARTS, Arterial Systolic BP |
| 3012410 | 20112-9 | LOINC | Tidal volume setting Ventilator | TV |
| 3025853 | 20140-0 | LOINC | Volume expired | VO2 |

We used case-sensitive row names to determine the measure of interest. For temperature, the strings 'temp', 'Temp', 'TEMP' had different associations in the flowsheet data. Although 'TEMP' contained integer values, the values were in the thousands which is out of range of a body temperature reading. Furthermore, the flowsheet group name was 'HD Machine Check'. Another row name, '(Retired) Temp' was a measure related to equipment, since the template name was 'Special Equipment' and the group name was 'Targeted Temperature Management'.

There were 130 row names where the row name contained the substring 'temp' and not the substring 'attempt'. Some were clearly not related to body temperature, such as 'Circuit Water Temperature', 'Device Set Attempt', 'Humidifier Temperature', 'Air Temp', 'Water Temperature'. 'Temp src' contained a text location of where the temperature was taken. There was uncertainty on the meaning of, and therefore whether to include 'Temp 2', 'Temp 3', 'Brain Temp', 'Patient Core Temperature' and others. For this initial version, we decided to include the following to represent body temperature: 'Temp', 'Temperature', 'Temp (in Celsius)', 'Tcore'.

Table 2: Top 30 temperature related row names and number of occurrences in Stanford's EHR.

| Temp | 43,163,193 |
|---|---|
| Temp src | 25,256,950 |
| Temp (in Celsius) | 12,952,769 |
| Tcore | 4,506,680 |
| Temp Source | 4,050,492 |
| In last 6 hours temperature < 36 C or > 38.3 C | 2,252,506 |
| RLE Temperature/Condition | 1,881,329 |
| LLE Temperature/Condition | 1,846,976 |
| LUE Temperature/Condition | 1,834,203 |
| Temperature RLE | 1,832,733 |
| Temperature LLE | 1,825,700 |
| Temperature RUE | 1,812,643 |
| Temperature LUE | 1,810,598 |
| RUE Temperature/Condition | 1,779,112 |
| Skin Temp, Distal to Site | 1,290,322 |
| Temp < 36 C (96.8 F) or >38.3 C (100.9 F) | 1,092,763 |
| Humidifier Temperature | 730,136 |
| Temperature | 706,614 |
| Temp 2 | 545,497 |
| Bed Set Temp | 494,263 |
| Humidifier Temperature (C) | 488,507 |
| Temperature (Blood - PA line) | 428,304 |
| Temperature Skin | 389,905 |
| Temperature greater than 35.5 C Ax. or 36 C oral within 45 minutes of discharge | 375,659 |
| Temp Control | 320,951 |
| Air Temp | 315,910 |
| Circuit Water Temperature | 168,098 |
| Temp Pacer Ventricular mA | 140,225 |
| Temperature > 37.8 C | 129,706 |
| Patient Core Temperature | 98,642 |

For pain level, we mapped the pain level at the first site, 'Pain Level - 1st Site', which was the most common and assumed to be the measurement of most interest to the researchers.  The LOINC code 38214-3 does not specify the location of pain with any granularity.  We decided to not add the 2nd, 3rd and 4th pain level sites because of questionable utility for this initial iteration but may consider including them in a future release.  The values of 'Pain Level - 1st Site' were integer values ranging from 0 to 10 with 10 being the highest level of pain. In line with only including numerical measured values, we also excluded pain measures where the values were 'Yes' or 'No.

For heart rate (see Table 3), the row name, 'Heart Rate', was clearly the dominant measure. We could, in the future, map other measures if requested by a researcher.

Table 3: The following table shows the top 22 'Heart Rate' related row names and number of occurrences in Stanford's EHR.

| | |
|---|---|
| Heart Rate | 84,587,071 |
| PEWS[1] Heart Rate Score | 10,377,491 |
| Post RT Treatment Heart Rate | 1,128,175 |
| Heart Rate Score | 915,259 |
| Fetal Heart Rate | 146,895 |
| Heart Rate Source | 71,183 |
| 2.Heart Rate greater than 110 beats/minute | 30,267 |
| Max Heart Rate (bpm) | 3,008 |
| Resting Heart Rate (bpm) | 3,007 |
| Heart Rate Recovery at 1 minute (bpm) | 3,005 |
| Measured Heart Rate Max | 2,443 |
| Fetal Heart Rate (FHR) | 1,845 |
| Fetal Heart Rate auscultated for more than 60 seconds | 1,124 |
| Heart Rate Used | 878 |
| 6 Minutes Heart Rate | 816 |
| Baseline Heart Rate | 810 |
| Heart Rate (b/min) | 337 |

---

[1] Paediatric Early Warning System

| | |
|---|---|
| In the last month, how many times a week do you usually do 30 minutes or more of moderate-intensity physical activity that increases your heart rate or makes you breathe harder than normal? (e.g., carrying light loads, bicyclingMu at a regular pace, or doubles tennis) | 282 |
| Fetal Heart Rate Monitoring | 201 |
| Beginning O2 Heart Rate | 200 |
| Ending O2 Heart Rate | 199 |
| Heart Rate > 100? | 75 |

The majority of the 28 row names required little data cleaning. The row names, 'Height', 'Weight', 'O2', 'Pulse', 'N2O', were matched exactly as is. We did not perform data cleaning on the measure values. For 'Heart Rate', the most common value was 2 as some flowsheet templates were designed to record the PEWS heart rate score yet abbreviated the label to 'Heart Rate'.  For temperatures, some templates recorded measurements in Celsius while others in Fahrenheit.  The measurement unit was not always available.  For a future release, we may consider using an algorithm to assign a unit of measurement to help researchers with this data cleaning process.

# Results

Using the methodology above, we bring in a large number of rows in OMOP measurements. Table 4 presents the total number of measurements brought in for 1,928,785 patients (~60% of total patients in the CDM). In Table 5, we break down the measurements by their LOINC code.

Table 4: Stanford flowsheet measures brought into OMOP representing 1,928,785 patients since 2008

| | |
|---|---:|
| Stanford Health Care (Adult) | 1,162,885,326 |
| Stanford Children's Health | 155,802,115 |

Table 5: Stanford OMOP measurement table counts from Flowsheets

| LOINC code | Concept Name | Count |
|---|---|---:|
| 8302-2 | Body height | 10,611,331 |
| 8277-6 | Body surface area | 11,922,167 |
| 8310-5 | Body temperature | 59,527,253 |
| 29463-7 | Body weight | 16,191,696 |
| 60985-9 | Central venous pressure (CVP) | 10,814,248 |
| 8462-4 | Diastolic blood pressure | 76,171,007 |
| 60802-6 | Dynamic plateau pressure | 35,028,360 |
| 35088-4 | Glasgow coma scale | 702,1807 |
| 8867-4 | Heart rate | 138,154,828 |
| 3151-8 | Inhaled oxygen flow rate | 112,276,832 |
| 60826-5 | Lung compliance | 28,362,146 |
| 60949-5 | Mean airway pressure | 45,958,305 |
| 8478-0 | Mean blood pressure | 50,271,563 |
| 60804-2 | Minimum alveolar concentration (MAC) | 93,510,451 |
| 33425-0 | Minute volume setting Ventilator | 39,363,808 |
| 60860-4 | Nitrous oxide [VFr/PPres] Gas delivery system | 96,786,321 |
| 19994-3 | Oxygen/Inspired gas setting [Volume Fraction] Ventilator | 55,709,852 |
| 76248-4 | PEEP Respiratory system --on ventilator | 36,562,214 |
| 38214-3 | Pain severity [Score] Visual analog score | 24,267,331 |
| 19931-5 | Peak inspiratory gas flow setting Ventilator | 37,143,121 |
| 8634-8 | Q-T interval | 11,823,427 |
| 8636-3 | Q-T interval corrected | 12,225,597 |
| 9279-1 | Respiratory rate | 103,608,124 |
| 60782-0 | Sevoflurane gas delivered during case [Volume] from Gas delivery system | 34,622,012 |
| 8480-6 | Systolic blood pressure | 82,305,997 |
| 20112-9 | Tidal volume setting Ventilator | 39,285,037 |
| 20140-0 | Volume expired | 33,157,095 |

Developing a clear, comprehensive and accurate data model of the health status of inpatients is challenging for a variety of reasons. Data captured by automated tools such as vital sign monitors is voluminous. Data captured in flowsheets, while a potentially rich source of detailed information on inpatient health status, is typically hand-entered and therefore prone to typographic variance and inconsistency that can be very difficult to interpret with computer algorithms. We balanced the competing needs for comprehensive coverage and accuracy by delivering all flowsheet entries verbatim in JSON format, augmented by selective mapping of high-value flowsheet types to controlled clinical vocabularies, specifically LOINC. In this way we have enabled our research community to build data cleaning algorithms as needed for the flowsheet entries not already mapped.

# Discussion

We have attempted to share the approaches we have adopted at Stanford to make the flowsheets data broadly usable by the research community. We trust that the approach we have presented is generalizable, even if Stanford ETL code is not directly reusable because of the lack of standardization of flowsheets in health systems.

We will continue to map other high value flowsheets in time. We encourage collaborative efforts with our research community that culminate in the contribution back into our OMOP of any newly developed mappings, and anticipate that over time our curated mappings of flowsheet values will become an invaluable source of insight into the health journeys of the inpatient population modeled in our dataset.

Finally, we mapped these flowsheets to OMOP measurements. In some cases, whether something should be part of measurements or observations is not completely clear. We hope to engage the OHDSI community in ongoing discussions so we can develop a common understanding.

# Acknowledgement

STARR suite (2008-), including the first generation data warehouse STRIDE, are made possible by Stanford School of Medicine Dean's Office. User training is supported by the National Center for Research Resources and the National Center for Advancing Translational Sciences, National Institutes of Health, through grant UL1 TR001085.

Using CRediT taxonomy (https://casrai.org/credit/), we present the contributing roles for our authors - Tina Seto (Software, Formal Analysis, Writing - original draft), Lillian Sung (Formal Analysis, Investigation, Methodology), Jose Posada (Methodology), Priyamvada Desai (Methodology, Project administration), Susan Weber (Writing - review & editing), Somalee Datta (Writing - review and editing).

We also thank the extended Stanford Medicine research community for their feedback (Stephen Pfohl, Jason Fries, Catherine Aftandilian) and Research IT team in bringing the software into production including Deepa Balraj (Software), Joseph Mesterhazy (Software), Wencheng Li (Software), Jaden Yang (Validation) and Nivedita Shenoy (Project administration).